\documentclass[aps,prl,showpacs,preprint,superscriptaddress]{revtex4}
\usepackage{graphicx}%
\usepackage{bm}%
\usepackage{amsmath,amssymb,amsfonts,wasysym,color}
\usepackage{dcolumn}%
\begin{document}
\title{ Casimir interactions in Ising strips with boundary fields: exact results. }
\author{Douglas~B. Abraham}
\affiliation{ Theoretical Physics, Department of Physics, University of Oxford, 1 Keble Road, Oxford OX1 3NP, United Kingdom}
\affiliation{Max-Planck-Institut f{\"u}r Metallforschung, Heisenbergstr.~3, D-70569 Stuttgart, Germany}
\author{Anna Macio\l ek}
\affiliation{Max-Planck-Institut f{\"u}r Metallforschung, Heisenbergstr.~3, D-70569 Stuttgart, Germany}
\affiliation{Institut f{\"u}r Theoretische und Angewandte Physik,
Universit{\"a}t Stuttgart, Pfaffenwaldring 57, D-70569 Stuttgart, Germany}
\affiliation{Institute of Physical Chemistry, Polish Academy of Sciences,
Department III, Kasprzaka 44/52, PL-01-224 Warsaw, Poland}
\date{\today}
\begin{abstract}
\vskip10pt
An exact statistical mechanical derivation  is given of the critical Casimir 
forces   for  Ising strips 
with arbitrary surface fields applied to edges. Our results show that 
the  strength as well as the sign of the force  can be controled  by varying 
the temperature or the  fields. An interpretation of the results is given  in terms of   a linked cluster expansion. This  suggests a  systematic  approach for  deriving the critical Casimir force 
which can be used in  more general models.
\end{abstract}
\pacs{05.50.+q, 64.60.an, 64.60.De, 64.60.fd, 68.35.Rh}
\maketitle

Casimir forces \cite{Casimir} arise in the quantum 
electrodynamics of geometrically restricted systems, for instance between two metal plates in vacuo, because the photon spectrum is modified; typically the force is attractive.
Fisher and de Gennes  \cite{FdG} proposed that analogous 
Casimir forces should arise in condensed matter systems
 near a second-order phase transition, the agent being 
thermally-excited 
fluctuations of the order parameter, e.g. of the density, rather then  of the  photon  field.
Of particular interest, both experimental and theoretical, is their scaling-theoretic prediction that such interactions should have a power law dependence on  distance in the critical scaling region. For example,  in spatial dimension $d$ for  walls separated by a distance $N$, the Casimir force per unit area  is ${\cal F}_{Cas} = -N^{-d}\vartheta(N/\xi)$, where $\xi$ is the bulk correlation length \cite{FdG,krech:99:0,Dbook}.
(All free energies and forces are expressed in units of $k_BT$.)
The tunability of critical
 Casimir interactions will be  crucial
 for  many applications in micro and nano systems, e.g. in colloids or in various micro or nano-electromechanical devices, in particular,  to be able to produce {\it repulsive}  interactions to counteract the omnipresent attractive Casimir quantum electrodynamical force.

If the system is confined to a film, one would expect
on intuitive grounds that the {\it geometrical} effect on the order parameter fluctuations
of   the strip boundaries,  would be to reduce the entropy; thereby  
establishing  a strictly {\it repulsive} force. 
This argument is incomplete because some energetic factors are neglected. 
We propose  an alternative view that admits attractive forces.
Let us examine the following approximate treatment  of a $2d$ Ising strip 
of finite width $N$  and  with free boundaries.
Following Privman and Fisher \cite{privman}, the low energy excitation of such a model are 
{\it domain walls}, each with free
energy $N\tau$ where $\tau$ is the incremental 
free energy per unit length   
for an interface perpendicular to the strip axis. A collection of these is then treated as a $1d$ Ising model with cyclic boundary conditions. A simple calculation
shows that the limiting incremental free energy, per unit length of strip, is $f^{\times}=-\log \left(1+e^{-N\tau}\right)$ and then the Casimir force, again per unit length, is 
$
{\cal F}_{Cas}=-\partial f^{\times}/\partial N=-\tau/(e^{N\tau}+1).
$
In the scaling limit $N\tau =x,  N\to \infty, \tau\to 0$, (\ref{eq:1}) becomes 
\begin{equation}
\label{eq:1}
N{\cal F}_{Cas}(x)\rightarrow -\frac{x}{e^{x}+1}.
\end{equation}
Notice that this force is {\it attractive}, but  the power of $N$ disagrees with
 \cite{FdG}.
What is missing in this formula? Firstly, there is no mention
of any thermally-excited intrinsic structure of the individual domain walls. This  could
perhaps  be included for the planar Ising model with free boundaries by calculating the partition function of a single interface connecting the edges with one end fixed, but the other free;   this merely reproduces (\ref{eq:1}).
Secondly,  the detailed interactions of these domain walls  must be quite complicated,  but they are evidently accounted for in
exact  calculations on  planar Ising strips.
The first known results \cite{ES} were for the strip with zero bulk field and either {\it free} boundary condition or {\it fixed} boundary spins, both the $++$ and $+-$ conditions. 
\begin{figure}
  \includegraphics[scale=0.3]{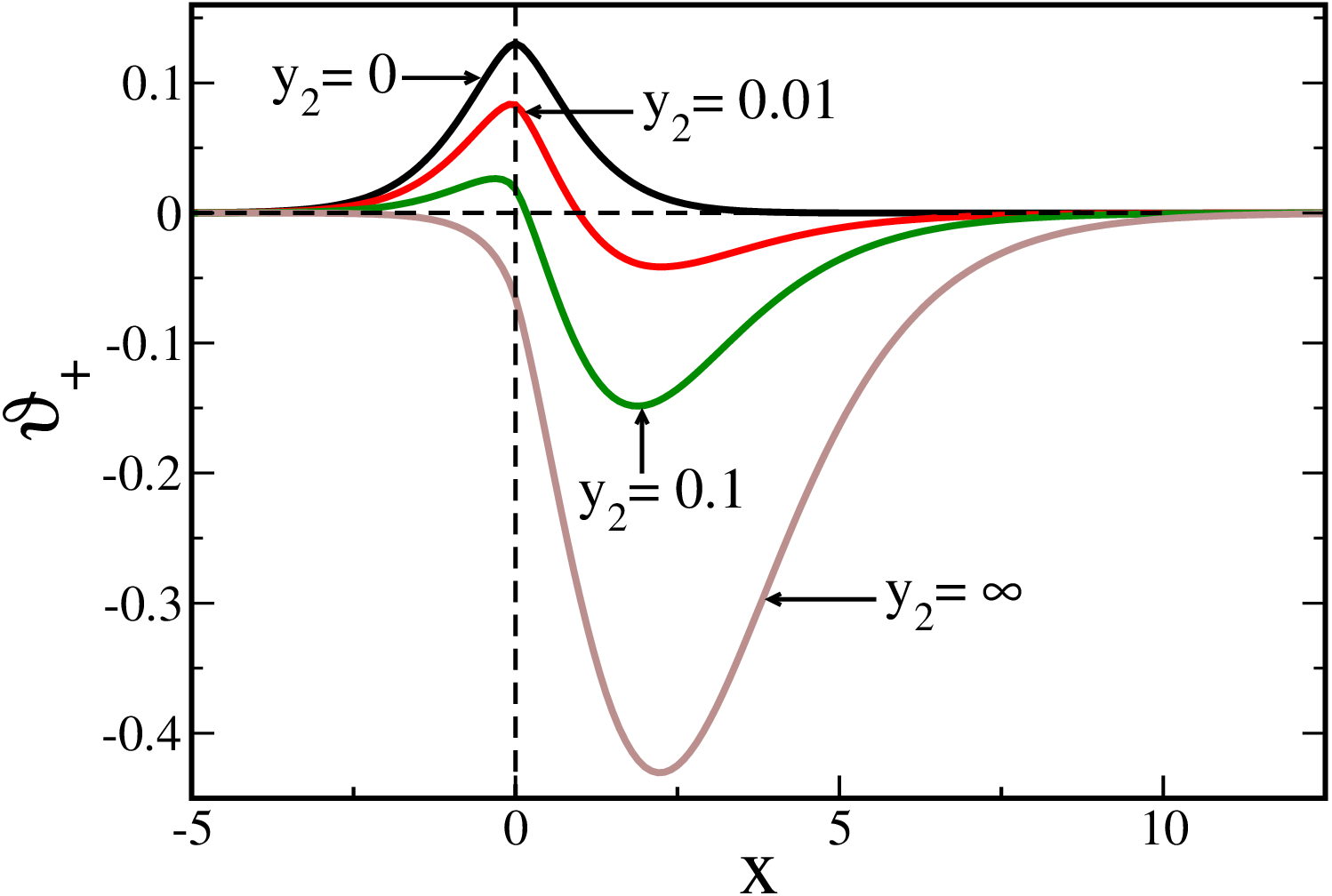}
 \caption{(color online) The scaling function $\vartheta_{+} (x,y_1,y_2)$ of the critical Casimir force (\ref{eq:10}) for the  isotropic lattice with $K_1=K_2$, $y_1=h_1^2N=\infty$ and for the several values of $y_2=h_2^2N$.}  \label{fig:1}
\end{figure}

The partition function for the {\it f}ree strip is \cite{DBMIT}
\begin{equation}
\label{eq:2}
Z_f=\prod_{\substack{ 0<\omega\le \pi \\ e^{iN\omega}=-1}}\left[\cosh N\gamma(\omega)+\sinh N\gamma(\omega) \cos \delta'(\omega)\right].
\end{equation}
The Onsager functions \cite{onsager}
$\gamma(\omega)$ and $\delta'(\omega)$ are given by 
$\cosh\gamma(\omega)=\cosh 2K_2\cosh 2K_1^*-\sinh 2K_2\sinh 2K_1^*\cos \omega$,
and
\begin{equation}
\label{eq:4}
e^{i\delta'(\omega)}=\left[\left(z-A\right)\left(z-B\right)/\left(Az-1\right)\left(Bz-1\right)\right]^{1/2}.
\end{equation}
where $z=e^{i\omega}$, $A=\exp2(K_1+K_2^{*})$ and $B=\exp2(K_1-K_2^{*})$.
 $K_{1}$ and $K_{2}$ are the nearest neighbour couplings (in units of $k_BT$) in $(0,1)$ and $(1,0)$ direction,
respectively.
$K^*$ is the dual coupling given by the involution
$\sinh(2K)\sinh(2K^*)=1$. 
The bulk free energy is given by extracting a factor $e^{N\gamma}/2$ from the argument of the logarithm leaving an incremental free energy. Removing an additional $N$-independent factor gives
\begin{equation}
\label{eq:5}
f^{\times}=\int_{-\pi}^{\pi}\frac{d\omega}{4\pi}\ln\left[1+(1/d'(\omega)e^{-2N\gamma(\omega)}\right],
\end{equation} 
where $d'(\omega)=(1+\cos\delta'(\omega))/(1-\cos\delta'(\omega))$.
The Casimir force per unit length is now 
\begin{eqnarray}
\label{eq:6}
{\cal F}_{Cas}(N,T)=
-\int_{-\pi}^{\pi}\frac{d\omega}{2\pi}  \frac{ \gamma(\omega)}{\left[ d'(\omega)e^{2N\gamma(\omega)}+1\right]}.
\end{eqnarray}
Comparing this with (\ref{eq:1}), we see that (\ref{eq:6}) also describes an attractive force, but there is also an integration, $\tau$ is replaced by
$\gamma(\omega)$ under the integral sign and we have $\exp 2N\gamma(\omega)$ rather than 
$\exp N\tau$ in the denominator. In addition, there is the prefactor $d'(\omega)$ multiplying this term. 
Taking the scaling limit of  (\ref{eq:6}) then recaptures 
the Fisher-de Gennes  law with the correct power.

The aims of this Letter are two-fold: firstly, we show how (\ref{eq:6}) is generalised to include non-zero surface fields, which allows us to control the sign of the Casimir force at will. Secondly, we interpret this result as a linked cluster expansion and then indicate why it might be appropriate for models more general than the planar Ising ferromagnet with nearest neighbour interactions.

\begin{figure}
   \includegraphics[scale=0.3]{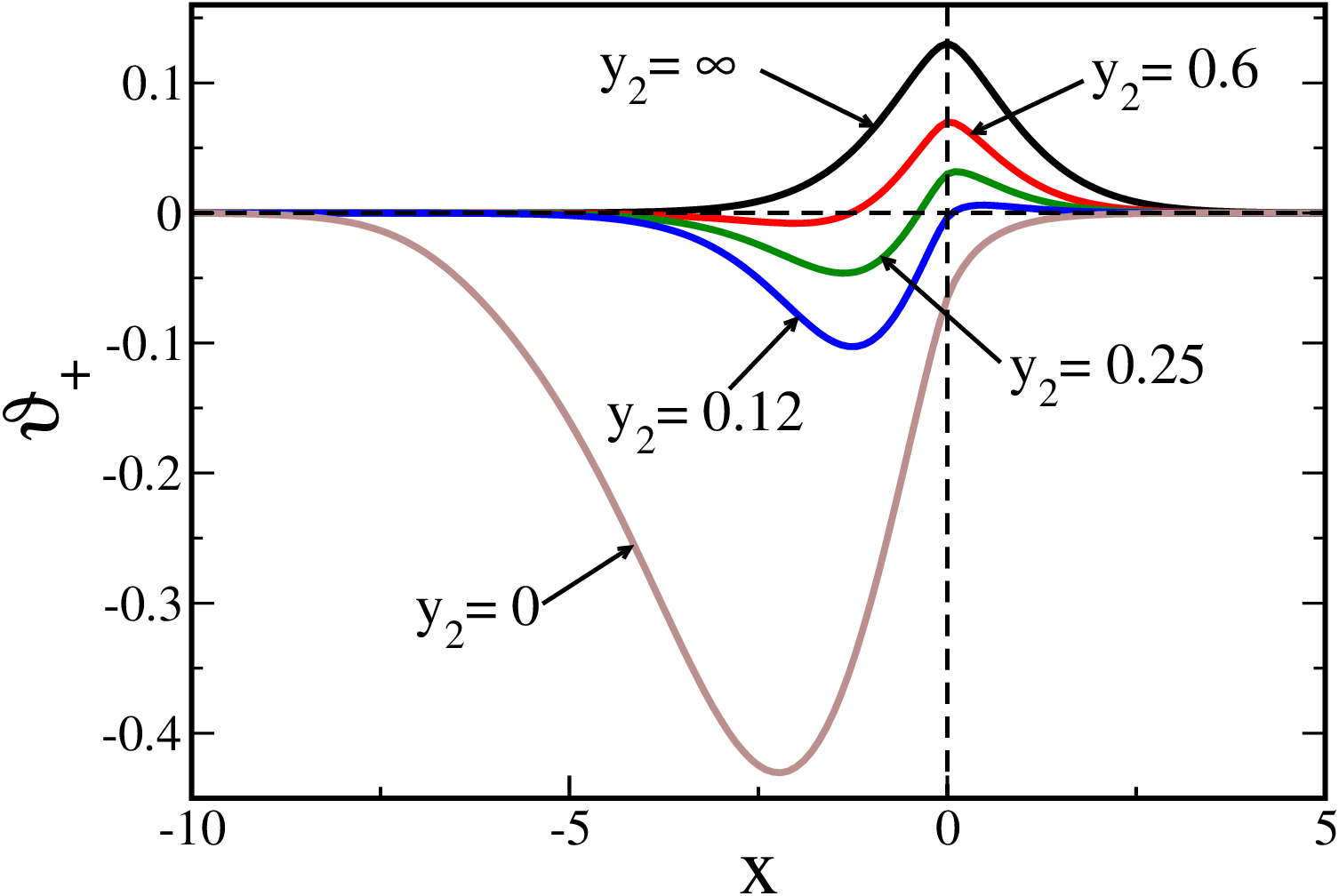}
 \caption{(color online) The scaling function $\vartheta_{+} (x,y_1,y_2)$ of the critical Casimir force (\ref{eq:10}) for the isotropic lattice with $K_1=K_2$  $y_1=h_1^2N=0$ and for the several values of $y_2=h_2^2N$.}
  \label{fig:2}
\end{figure}

A formalisation   of  the technique of  Schultz, Mattis and Lieb \cite{LMS} allows us 
to calculate the partition function of a cylindrical lattice  with circumference $M$,  height $N$, with its axis in $(0,1)$
direction, and end fields  $h_j > 0, j=1,2$  in a straightforward way. 
The fields are introduced by taking a free-edged cyclic strip and adding an extra ring of spins at each end; these spins are forced to take the value $+1$. These fixed spins are then coupled to the free lattice by bonds of  strength $h_1$  at the bottom and $h_2$ at the top. A different technique will  be needed if $h_1h_2<0$, as will be seen.
The Casimir force per unit length in the $(1,0)$ direction as $M\to \infty$ is obtained
from the incremental free energy by taking the derivative in respect to $N$:
\begin{eqnarray}
\label{eq:8}
{\cal F}_{Cas}(h_1,h_2,N,T)= -\int_{-\pi}^{\pi}\frac{d\omega}{2\pi} \frac{\gamma(\omega)} {1+\frac{d^{*}(\omega)}{A_1A_2}e^{2N\gamma(\omega)}}
\end{eqnarray} 
where $d^{*}(\omega)= (1+\cos \delta^*(\omega))/(1-\cos \delta^*(\omega))$ and
$A_j=\left(e^{-\gamma(\omega)}-w_j\right)/\left(e^{\gamma(\omega)}-w_j\right),\quad  j=1,2$. 
$
e^{i\delta^*(\omega)}
$
is given by (\ref{eq:4}) with $B$ replaced by $B^{-1}$.
The values $w_1$ and $w_2$ which are the wetting parameters for the force are given by \cite{DBA80}
$
w_j=e^{2K_2}\left( \cosh 2K_1-\cosh 2h_j\right)/\sinh 2K_1.
$
It is crucial to note that $A_j(\omega)$ can take  both positive and negative values; this is why either sign of the Casimir force is possible in principle.
In the scaling limit, $N\to \infty, \gamma(0)=K_2-K_1^{*} \to 0$ such that  $x=N\gamma(0)sgn(T-T_c)$
is fixed (as $t\equiv (T-T_c)/T_c\to 0, K_2-K_1^{*}\simeq -4K_ct$) and $N\omega =u$ leads to
${\cal F}_{Cas}(h_1,h_2,N,T)=N^{-2}\vartheta_{+}(\underline{ r})$, with
\begin{equation}
\label{eq:10}
\vartheta_{+}(\underline{ r})=-\frac{1}{\pi}\int_0^{\infty} \frac{du\lambda (x,u)}{\frac{X^-(0)}{X^+(0)}\frac{X^+_1X^+_2}{X^-_1X^-_2}e^{2\lambda(x,u)}+1},
\end{equation}
where $\underline{ r}=(x,y_1,y_2)$ with $y_j=h_j^2N$,  $\lambda (x,u)=\sqrt{x^2+u^2}$ and $X^{\pm}(y)=\lambda(x,u)\mp\left(x-2e^{2K_c}y\right)$,  $X^{\pm}_j= X^{\pm}(y_j), j=1,2$. For $T<T_c$, $\gamma(0)$ is the surface tension in the
$(0,1)$ direction of the $2d$ Ising model; it  is the inverse correlation length for $T>T_c$. At $x=0$
(\ref{eq:10}) reduces to the universal Casimir amplitude, which equals $-\pi/48$
for both $h_j=0$ and  $h_j=\infty$. Interesting examples of the scaling function $\vartheta_{+}(\underline{r})$, which demonstrate that the critical Casimir forces can switch from attration to repulsion by varying the temperature, are shown in Figs.~\ref{fig:1} and ~\ref{fig:2}. They were evaluated numerically from (\ref{eq:10}) for several choices of the scaling variables $y_{1,2}$.

\begin{figure}
   \includegraphics[scale=0.3]{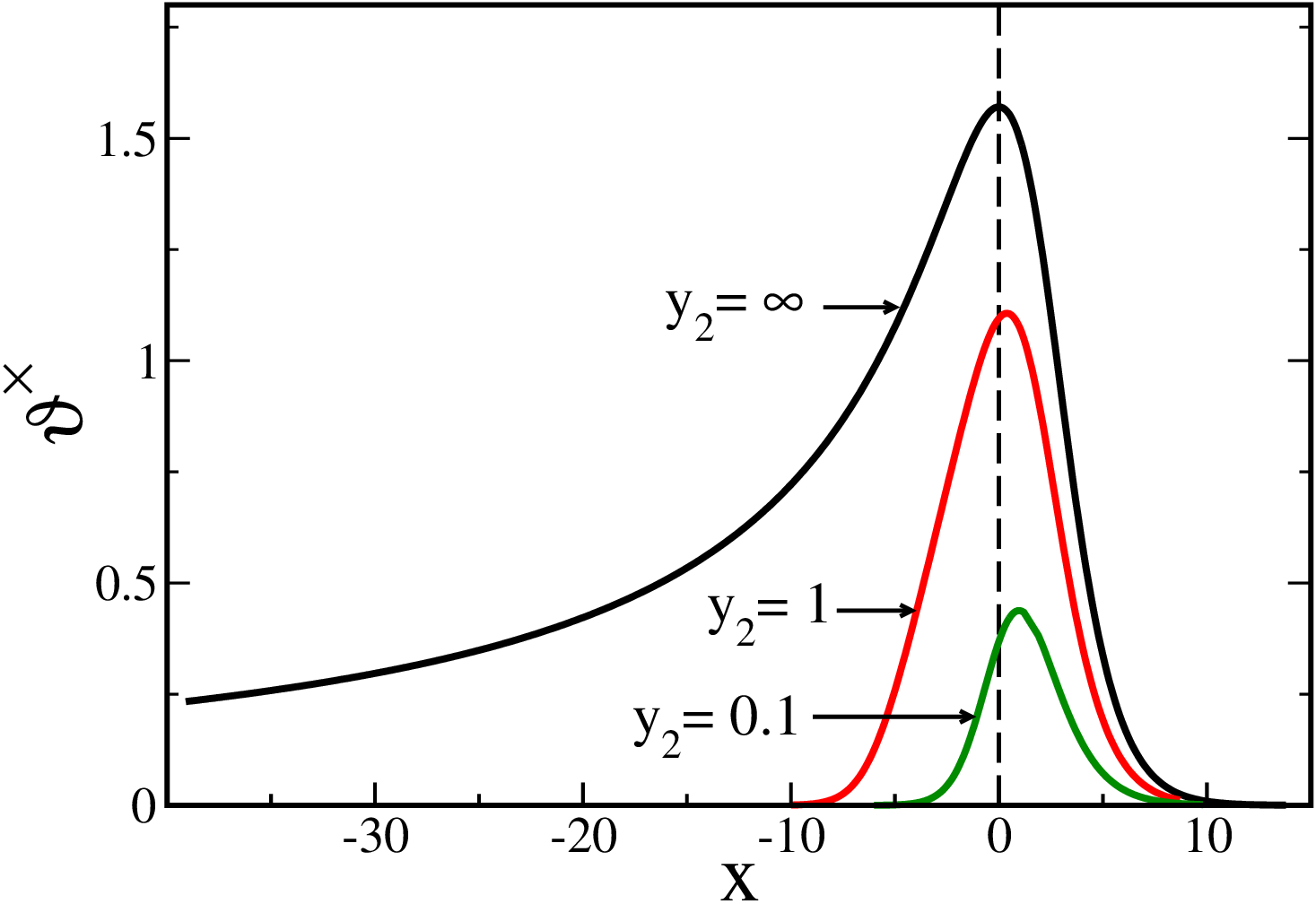}
 \caption{(color online) The excess  scaling function $\vartheta^{\times}(x,y_1,y_2)$ of the critical Casimir force (\ref{eq:13}) for the isotropic lattice with $K_1=K_2$  $y_1=h_1^2N=\infty$ and for the several values of $y_2=h_2^2N$.}
  \label{fig:3}
\end{figure}

\begin{figure}
   \includegraphics[scale=0.3]{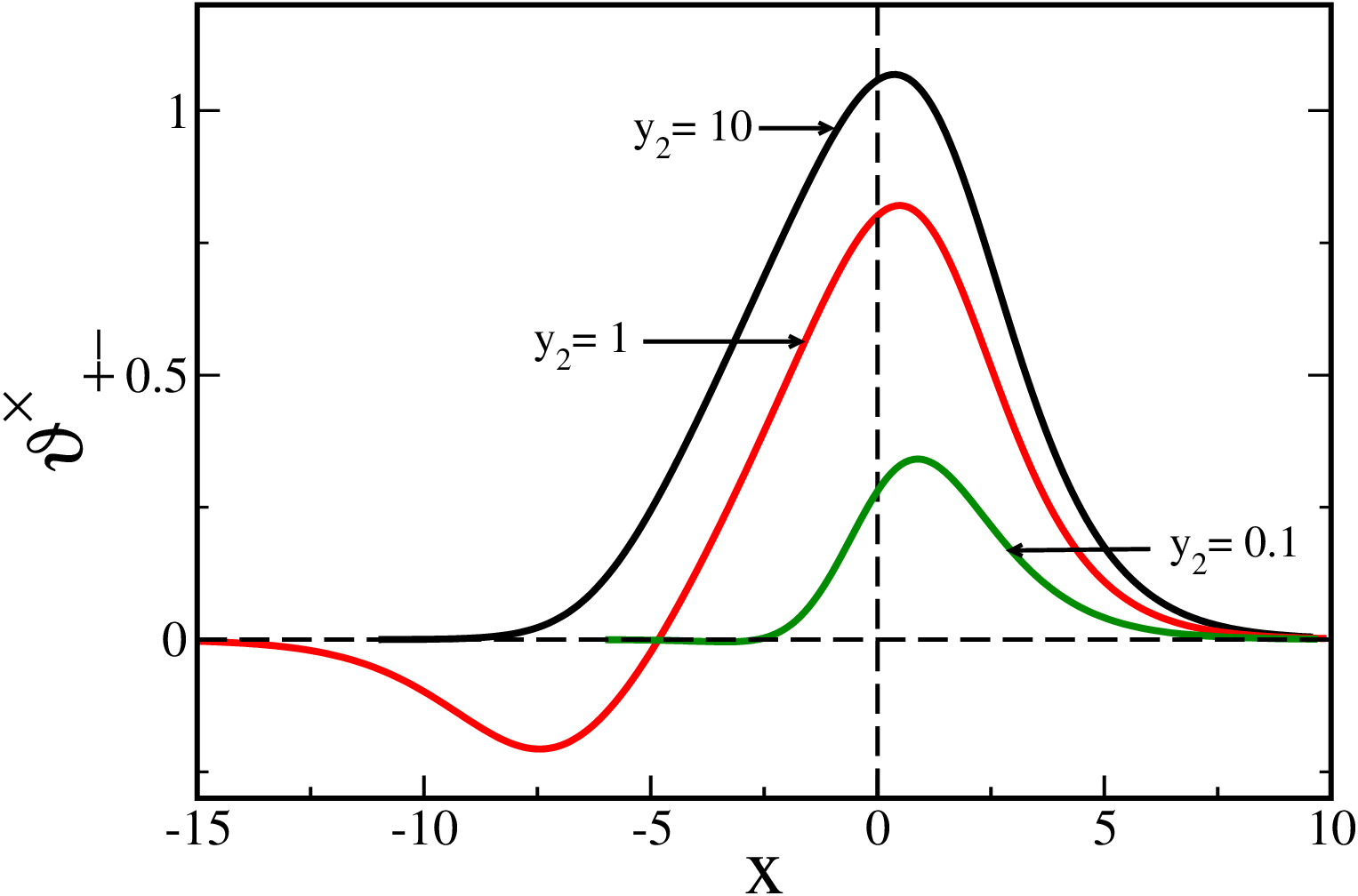}
 \caption{(color online) The excess  scaling function $\vartheta^{\times} (x,y_1,y_2)$ of the critical Casimir force (\ref{eq:13}) for the isotropic lattice with $K_1=K_2$  $y_1=h_1^2N=1$ and for the several values of $y_2=h_2^2N$. }
  \label{fig:4}
\end{figure}

The case with $h_1h_2<0$ can be approached from that with $h_1h_2>0$ by reversing the end spins between $x=1$ and $x=s+1$ on one face of the cylinder thereby creating an interface with terminations in the same face.  This is followed by taking the limit as $M\to \infty$, as before. With $h_j>0, j=1,2$, we find  the ratio of partition functions for strips with and without the interface to be:
\begin{eqnarray}
\label{eq:11}
e^{-sf^{\times}}= 
\int_{-\pi}^{\pi}\frac{d\omega}{2\pi}\frac{ i e^{is\omega} \tan (\delta^*/2)\left(B^+-B^-A_2e^{-2N\gamma(\omega)}\right)}{1+A_1A_2e^{-2N\gamma(\omega)}}
\end{eqnarray}
where $B^{\pm}=\left(e^{\pm\gamma(\omega)}-e^{-4K_2}w_1\right)/(e^{\gamma(\omega)}-w_1)$. 
We are interested in the limit $s\to \infty$ of the rhs of (\ref{eq:11}) per unit length. The asymptotics 
for large $s$ is dominated by the nearest singularity to the real axis in the strip
$-\pi < \omega  \le \pi$. The branch cuts associated with $\sinh \gamma(\omega)$ do not occur and
poles are simple  zeros of the denominator of (\ref{eq:11}). Fortunately, 
the problem can be related to the diagonalisation problem of the  transfer matrix in the direction $(1,0)$ \cite{AM} (here we are transferring in $(0,1)$ direction) by  looking  for the solution in the variable $k$  such that 
 $\omega=i{\hat\gamma}(k)$  and $\lim_{\epsilon \to 0^+}\gamma(i{\hat\gamma(k)\pm\epsilon})=\pm ik$, where the function ${\hat\gamma}(k)$ is defined as the
Onsager function, but with $K_1$ and $K_2$ interchanged.
Then finding zeros of  the denominator of (\ref{eq:11}) becames equivalent to solving the spectrum discretisation 
condition for the strip transfer matrix in the $(1,0)$ direction,
which was  studied in detail in Ref.~\cite{AM}:
\begin{equation}
\label{eq:12}
e^{2i(N+1)k}=e^{2i{\hat\delta}'(k)}\frac{e^{ik}w_1-1}{e^{ik}- w_1}\frac{e^{ik}w_2-1}{e^{ik}-w_2}.
\end{equation}
$e^{i{\hat\delta}'(k)}$ is obtained from $e^{i\delta'(k)}$  by interchanging $K_1$ and  $K_2$.
In the  scaling limit $N\to \infty$, $K_2-K_1^*\to 0$ and 
$N(K_2-K_1^*)=x$, we find  $f^{\times}=(1/N)\lambda(x,u_0)$. Hence  the solution for the Casimir scaling function   $\vartheta_{-}(\underline{ r})=\vartheta^{\times}(\underline{ r})+\vartheta_{+}(\underline{r})$ has the  implicit form with
\begin{equation}
\label{eq:13}
\vartheta^{\times}(\underline{ r})=\frac{u_0^2-u_0{\underline r}\cdot \underline{ \nabla}u_0}{N^2\lambda(x,u_0)},
\end{equation}
where   $u_0({\underline r})$  solves
 the quantisation condition (\ref{eq:12}) in the scaling limit  
\begin{equation}
\label{eq:14}
e^{2iu}=-\frac{Z^+(0)}{Z^-(0)}\frac{Z_1^-Z_2^-}{Z_1^+Z_2^+},
\end{equation}
where $Z^{\pm}_j= Z^{\pm}(y_j), j=1,2$ is derived from $X_{\pm}(y)$ by replacing  $\lambda(x,u)$  by $iu$.
The derivatives of $u_0$ can be calculated straightforwardly from (\ref{eq:14}).
In Figs.~\ref{fig:3} and \ref{fig:4} we plot
$\vartheta^{\times}$ as a function of $x$ evaluated numerically for some choices of the
scaling variables $y_1$ and $y_2$. 
Our results for the special case of $y_1=y_2$ agree with those reported in Ref.~\cite{napiorkowski}; the change of   sign of the  scaling function $\vartheta^{\times}$ is associated with the  localisation-delocalisation transition  \cite{parryevans}. This feature remains for a slightly broken symmetry, i.e., for $y_1\approx y_2$ and $y_{1,2}$ small. For strongly asymmetric strips the excess scaling function of the critical Casimir force is always positive.

We now interpret (\ref{eq:8}) in terms of statistical mechanical ideas. Expanding the integrand gives 
\begin{equation}
\label{eq:15}
f^{\times}=\sum_{n=1}^{\infty} \frac{(-1)^{n}}{n} \int_{-\pi}^{\pi} \frac{d\omega}{4\pi} e^{-2Nn\gamma(\omega)} \left(C_1C_2\right)^n.
\end{equation}
where $C_j(\omega)=\frac{A_j^-}{A_j^+}\tan \frac{\delta^*(\omega)}{2}, j=1,2$.
Although this is not immediately apparent, this is in fact a linked cluster expansion as we now show. Equation (\ref{eq:15}) can be understood by going back to the partition function formula in terms of a transfer matrix $V$
\begin{equation}
\label{eq:16}
Z=\langle b_1 \mid V^N \mid b_2 \rangle
\end{equation}
where $\mid b_j\rangle$ describes the edge state with field $h_j, j=1,2$. 
Instead of using the  Schultz, Mattis and Lieb technology, this can be developed by expanding with basis of eigenvectors of $V$ giving 
\begin{equation}
\label{eq:17}
\frac{Z}{\Lambda_{max}^N}= \sum_{n=0}^{\infty}\sum_{(\omega)_{2n}}\frac{e^{-N\sum_{j=1}^{2n}\gamma (\omega_j)}}{(2n)!}\langle b_1\mid(\omega)_{2n}\rangle\langle(\omega)_{2n}\mid b_2\rangle,
\end{equation}
where $\mid (\omega)_{2n}\rangle $ denotes a $2n$-fermion eigenstate of $V$ \cite{LMS}.
This would certainly not be the chosen way of obtaining (\ref{eq:8}), since we would need to evaluate the matrix elements $\langle b_1\mid(\omega)_{2n}\rangle, \langle(\omega)_{2n}\mid b_2\rangle$; this has been done with some effort and the result is typically Wick-theoretic in form:
\begin{equation}
\label{eq:18}
\langle b_j\mid (\omega)_{2n}\rangle = \sum_{m=2}^{2n} (-1)^m f_j(\omega_1,\omega_m)\langle b_j\mid \Delta_{1m}(\omega)_{2n}\rangle
\end{equation}
where $\Delta_{1m}(\omega)_{2n}=(\omega_2,\ldots \omega_{m-1},\omega_{m+1}\ldots \omega_{2n})$ and 
$
f_j(\omega_1,\omega_2)=iC_j(\omega)\delta_{\omega_1,-\omega_2}
$
is the contraction function or, alternatively, a scattering matrix element for a pair of fermions off the wall described by $\mid b_j \rangle$. Notice that since the $A^{\pm}_j(\omega)$ are even, but $\tan \frac{1}{2}\delta^*(\omega)$ is odd, the contraction is antisymmetric as it should be for fermions. Thus, (\ref{eq:18}) is a Pfaffian \cite{Pf}.
The graphical representation of (\ref{eq:17}) and (\ref{eq:18}) is discussed in \cite{DBA78}. For each $n$ we have a weighted sum of disjoint loops, each having an even number of vertices, the vertex weight $C_j(\omega)$ and the Kronecker delta edge weight.
The occurrence of the Kronecker delta  in the contraction function is mandated by translational symmetry. 
Thus, the multiple sum for each loop becomes just a single sum on implementing the deltas.
Asymptotically  as $M\to\infty$, each such sum is to leading order $M$ times a single integral. We can now apply the linked cluster theorem  to exponentiate (\ref{eq:17}). Eq.~(\ref{eq:15}) is recaptured, for the {\it excess free energy} per unit length in the $(1,0)$ direction since the factor of $1/n$ in (\ref{eq:15}) comes directly from a symmetry number argument \cite{uhlenbeck}.
 Each term is then to be thought of as a weight of a ''loop'' with $2n$ vertices. The loop is reflected $n$ times off the upper boundary and $n$ times off the lower boundary, with ''momentum''  conservation at each reflection; thus $n$ maybe thought of as a topological quantum number.  Starting from 
(\ref{eq:16}), (\ref{eq:17}) and (\ref{eq:18}),  we have re-derived  (\ref{eq:15}), in a way 
which allows us to  identify 
the multiplier of $\exp 2N\gamma(\omega)$ 
in (\ref{eq:8}) as a product of two scattering matrix elements, one from each edge. Clearly $\gamma(\omega)$ in (\ref{eq:17}) is a Fermion energy. Thus we have a complete {\it intuitive} understanding of (\ref{eq:8}). We can take the scaling limit either in (\ref{eq:15}) or  (\ref{eq:8}) (as we have already done) with the same outcome. This procedure even converges after taking $T\to T_c$ in either (\ref{eq:15}) or  (\ref{eq:8}),
since then $\gamma(\omega) \propto \mid \omega \mid$.

Two approximation schemes are in order. Firstly, we could consider how well partial sums of the virial series (\ref{eq:15}) approximate the exact result, so that we can assess the contribution of the different reflection number sectors to the result. Secondly, in the scaling limit the one-particle energy should be universal. The same is not likely to be true for the scattering matrix elements. Correlation droplet theory \cite{DBA83} provides an approximate method for calculating them and thus for extending the scope of our results.

In this Letter, we have described new, general, exact results for the critical Casimir force in a planar, rectangular Ising ferromagnet with applied fields $h_1$ and $h_2$ on the edges.
Each field can have arbitrary sign and magnitude. Both with $h_1h_2>0$ and with $h_1h_2<0$, we show that the force can be {\it attractive} or {\it repulsive}, according to the tuning of the parameters. The compensation of attractive, {\it quantum} van der Waals forces which  this will allow has implications which may well prove crucial for applications.
Mean field calculations are in qualitative agreement with our results \cite{mohry}. 
There are also related results from the continuum model ~\cite{diehl}.
We also interpret the representation of the Casimir force as in (\ref{eq:6}), (\ref{eq:8})
and (\ref{eq:15}) in terms of the {\it linked cluster} expansion. This suggest an associated droplet picture which enhances the original finite size scaling ideas of Privman and Fisher 
\cite{privman} in this context; this will also give new, systematic approximations for calculating 
critical Casimir forces in planar systems and perhaps even in $d=3$.
Our results can be directly applied to e.g. $2d$ binary fluid membranes with protein inclusions
close to the demixing point \cite{membranes}.

DBA acknowledges Max-Planck-Gesellschaft for hospitality.

\end{document}